# *Ab initio* Study on Lithium Anode Interface Instability and Stabilization of Superionic Li$_3$InCl$_6$ and Li$_6$PS$_5$Cl Solid Electrolytes

Cheng-Man Wang[1], Chao-Hsiang Hsu[1], Jing-Sen Yang, Ping-Chun Tsai[*]

Department of Mechanical Engineering, National Taiwan University of Science and Technology, Taipei city 106335, Taiwan

**Abstract**

Emerging superionic conductors Li$_3$InCl$_6$ (LIC) and Li$_6$PS$_5$Cl (LPSC) are very promising for solid-state electrolytes (SSEs) in all-solid-state lithium batteries (ASSLBs). However, unstable lithium-anode interfaces in LIC and LPSC have been observed through experiments and *ab initio* calculations, while the interphases formed in determining interfacial stability remain unclear. In this study, we investigate the *ab initio* stability of LIC-Li and LPSC-Li interfaces and interphases. Our *ab initio* calculations revealed that both interfaces are not chemically or electrochemically thermodynamically stable. Interestingly, the LPSC-Li has a stable interphase, but the LIC-Li doesn't. Moreover, the interlayers were systematically evaluated for the LIC-Li and LPSC-Li interfaces, and the desired interlayer materials are clarified for the two interfaces. This *ab initio* understanding of the interfaces, interphases and interlayers would help the development of a variety of stable interfaces in ASSLBs.

**Keywords**

**Introduction**

With the increasing demands for electric vehicles and grid storage systems, all-solid-state lithium batteries (ASSLBs) become promising energy-storage technologies because of their potentially high safety by using non-flammable solid-state electrolytes (SSEs). In addition, SSEs could physically inhibit the formation or growth of lithium dendrites, which allows the practical use of lithium metal to achieve the theoretical energy density of the negative electrode materials. The key performance metrics for SSEs are ionic conductivity and stability. A variety of superionic inorganic SSEs have been widely investigated, such as oxides (LIPON and LLZO), antiperovskites[1-3], sulfides (LGPS, LPS and LPSC) and halides (LIC)[4,5]. More recently, two very promising families of SSEs are metal halides and argyrodite sulfides. $Li_3InCl_6$ (LIC)[6] and $Li_6PS_5Cl$ (LPSC)[7] represent the two families. LIC has ionic conductivity ~1 mS/cm and a high oxidation potential of 4.3 V, but it experiences poor stability against Li-anode. Similarly, the LPSC family can achieve excellent ionic conductivity up to ~10 mS/cm but also show poor stability against Li-anode.

Li-anode interface instabilities of LIC and LPSC result from not only reduction decompositions of LIC and LPSC themselves but also interfacial chemical reactions with a Li-anode. LIC and LPSC both show a relatively high reduction potential of ~2 V versus $Li/Li^+$ [8,9], and they are likely to be reduced under ASSLB charging conditions. The interfacial chemical instability of LIC and LPSC with a Li-anode has been seen as one main cause for a huge increase in the interfacial resistance.[10-12] Side products at the anode interface have been observed by various experimental characterizations such as X-ray diffraction (XRD), X-ray photo-electron spectroscopy (XPS), and Raman spectroscopy.[12-14] It can be found that the LIC and LPSC both show poor electrochemical and chemical stability against a Li-anode. However, no satisfactory *ab initio* study has emerged that examines the Li-anode interface instabilities of LIC and LPSC with a Li-anode and solid-electrolyte interphase (SEI) properties.[15-19] This motivates us to use *ab initio* calculations to understand the interfacial stabilities of LIC and LPSC with a pure Li-anode and find out a protective interlayer to stabilize the Li-anode interface.

Introducing an interlayer is one effective approach to mitigate the anode interfacial degradation of ASSLBs, aiming to create a stable and protective layer for the SSE-anode interface. To the best of our knowledge, 29 kinds of ceramic materials of oxides, nitrides, sulfides and halides and 15 kinds of alloys have been reported to enhance Li-anode interface stability and increase the cycle life of ASSLBs.[20-66] However, the performance of the interlayers largely depends on the different processing technologies and parameters used, and interlayer coating experiments are often time-consuming and labor-intensive. Therefore, it is challenging to identify the most effective interlayers among the reported effective interlayers. High-throughput *ab initio* screening methods have been developed for anode coating materials in ASSLBs.[67-72]

In this study, we applied *ab initio* calculations to investigate the Li-anode interface stability of the LIC and LPSC solid-state electrolytes. The phase diagrams were used to calculate the thermodynamic phase stability, electrochemical stability and chemical stability of the LIC-Li and LPSC-Li interfaces, and their reaction products were evaluated to understand the stability of the solid-electrolyte interphase (SEI). To stabilize the LIC-Li and LPSC-Li interfaces, we systematically screened the 44 kinds and 100 compositions of the reported effective ceramics and alloy interlayers by evaluating their phase stability, chemical stability, electrochemical stability, electronic band gap and Li migration barrier. Finally, the best promising anode interlayer materials were identified for stabilizing the LIC-Li and LPSC-Li interfaces.

## Methods

### *Ab initio* Calculations

The settings and energies for most materials in this study are derived from the Materials Project (MP) database[73], many of which have been tested and are consistent with the calculations presented in **Table S1**. For materials or material properties not available in the MP database, density functional theory (DFT) calculations were performed using the Vienna Ab Initio Simulation Package (VASP). The electron exchange-correlation interactions were modeled using the generalized gradient approximation (GGA) and Hubbard U corrections (GGA + U), parameterized by the Perdew–Burke–Ernzerhof (PBE) functional. Projector augmented wave (PAW) pseudopotentials. All parameters of DFT calculations, such as the plane wave energy cut-off and k-points density, were consistent with the parameters used in the Materials Project (MP).[74]

### Phase Diagram and Phase Stability

The convex energy hull of phase and competing decomposition phases for the materials studied were computed using a MATLAB code developed by the authors in a previous study.[75] All possible phases comprising the constituent elements were obtained from phase diagrams in the Materials Project (MP) database.[73] The phase diagram showing only the lowest energy phases was constructed by the convex hull of the energies of all phases within the Li-In-Cl and Li-P-S-Cl compositional space. Competing decomposition phases were determined by considering all stable phases with the constituent elements under thermodynamic conditions. The convex energy hull ($E_{hull}$) of a phase can be calculated by Equation 1:

$$\Delta E_{hull} = E(c_{\text{phase}}) - E[c_{eq}(c_{\text{phase}})] \quad (1)$$

where the $c_{phase}$ is the composition of the phase, the $c_{eq}(c_{phase})$ is the phase equilibria of the $c_{phase}$, $E(c_{phase})$ and $E(c_{eq}(c_{phase}))$ represent the *ab initio* calculated energy in a unit of electron-volt per atom of the phase and its competing decomposition phases (phase equilibria), respectively.

**Electrochemical Stability and Potentials**

The redox reaction between SSE and Li-anode involves the exchange of electrons and Li ions, thus SSE behaves as an open system for lithium. The electrochemical (de-)lithiation of SSE can be represented as a SSE-Li pseudo-binary tie line in the phase diagram. Along this tie line, the SSE-Li phase equilibria can be established. The redox reaction energy ($\Delta E_{redox}$) can be calculated by Equation 2:

$$\Delta E_{redox}(\text{phase}, x) = E\{c_{eq}[(1-x)c_{phase} + xc_{Li}]\} - (1-x)E(c_{phase}) - xE(c_{Li}) \quad (2)$$

where the $c_{Li}$ is the composition of Li, $x$ and $(1-x)$ are the molar fractions of Li and phase in a SSE-Li system, respectively. The equilibrium potential ($\phi_{eq}$, V) referred to Li/Li$^+$ can be determined by the redox reaction energy ($\Delta E_{redox}$, eV) and the molar fraction of Li ($x$) during (de-)lithiation to SSE. Which can be calculated by Equation 3:

$$\phi_{eq} = \Delta E_{redox} / -x \quad (3)$$

**Interfacial Chemical Stability**

We employed a methodology developed by Ceder et al. to calculate the chemical stability between two reactants[76], which is related to bulk thermodynamics but neglects the kinetics of interfacial layer formation. This method identifies reaction products formed with the largest driving force when two reactant materials *A* and *B* are combined. The mutual chemical reaction energy, $\Delta E_{chem}$, is calculated by subtracting the *ab initio* calculated energies of reactants *A* and *B* from the product ground-state energy at the *A*/*B* interface at varying mixing atomic fraction *x* (1−*x* for reactant *A*, *x* for reactant *B*), as described by Equation 4:

$$\Delta E_{chem} = E\{c_{eq}[(1-x)c_A + xc_B]\} - (1-x)E(c_A) - xE(c_B) \qquad (4)$$

**Li Migration Barrier**

The migration barriers of Li$^+$ ions were determined using the climbing-image nudged elastic band (CI-NEB) method, which finds the minimum energy paths and saddle points for Li$^+$ ion migration, as implemented in the VASP with the Transition State Tools for VASP (VTST).[77,78] We constructed 3 × 3 × 3 supercell models and optimized the lattice constants for each structure. The supercells consist of 216 atoms for LiX (X = F, Cl, Br and I) with the rocksalt structure and 324 atoms for Li$_2$S with the antifluorite structure. The plane-wave cutoff energy used was 520 eV. A single Li vacancy in the supercell was created for an available diffusion site of Li$^+$ ions, and then atomic positions were optimized until residual forces were less than 0.02 eV/Å. Five images were set along each diffusion path to represent the ion migration trajectory. The convergence criterion for atomic forces was maintained at 0.02 eV/Å for each image. Numerical integration over the Brillouin zone was performed by using Monkhorst-Pack k-point grids of 2 × 2 × 2 for all compounds.

## Results and discussion

### Stability of Materials and Interfaces

The two state-of-the-art superionic halide $Li_3InCl_6$ (LIC) and argyrodite $Li_6PS_5Cl$ (LPSC) were studied. **Table 1** shows the space group, phase stability and electrochemical stability of LIC and LPSC. First of all, phase stability is an essential stability property of materials. *Ab initio* convex energy hull ($E_{hull}$) of a phase against the competing phases can be employed to quantify the phase stability of LIC and LPSC. The competing phases of LIC and LPSC can be determined by considering all stable phases composed of the constituent elements in the Li-In-Cl and Li-P-S-Cl phase diagrams shown in **Figures 1 (a) and 1 (b)**, respectively. The LIC phase has negative convex energy hull of -28 meV/atom with respect to a combination of its competing phases $InCl_3$ and LiCl, as $Li_3InCl_6 \rightarrow InCl_3 + 3\ LiCl$ ($\Delta E_{hull} = $ -28 meV/atom). This indicates that the LIC phase is thermodynamically stable. However, the LPSC phase exhibits little positive convex energy hull of 19 meV/atom against its competing phases $Li_3PS_4$, $Li_2S$, LiCl, which suggests that the LPSC phase could be thermodynamically metastable although LPSC can be synthesized.

**Table 1** Space group, calculated phase stability, and electrochemical stability of the halide $Li_3InCl_6$ (LIC) and argyrodite $Li_6PS_5Cl$ (LPSC) electrolytes, as well as the Li-anode

|  | Composition | Space group | Convex energy hull (meV/atom) | Decomposition products | Redox potential (V vs. Li/Li$^+$) | Redox products |
|---|---|---|---|---|---|---|
| Electrolyte | $Li_3InCl_6$ (LIC) | *C*2 | -28 | $InCl_3$, LiCl | Ox. 4.34 Red. 2.47 | Ox. $InCl_3$, $Cl_2$ Red. $InCl_2$, LiCl |
|  | $Li_6PS_5Cl$ (LPSC) | *F*-43*m* | 19 | $Li_3PS_4$, $Li_2S$, LiCl | Ox. 2.00 Red. 1.74 | Ox. $Li_3PS_4$, $LiS_4$, LiCl Red. LiCl, $Li_2S$, P |
| Anode | Li | *R*-3*m* | 0 | N/A | Ox. 0.00 Red. 0.00 | Ox. Li$^+$ Red. Li |

The equilibrium potential of the LIC and LPSC phases can be calculated using the phase equilibria of the Li-In-Cl and Li-P-S-Cl compositional spaces upon (de-)lithiation. Using the LIC as an example in **Figure 1 (c)** and **Table S2**, the reduction reaction of LIC occurs with an electrochemical lithiation to LiCl and $InCl_2$ at a voltage below $\phi_{eq}$ = 2.47 V, as $Li_3InCl_6$ + Li → 4 LiCl + $InCl_2$. Upon more lithiation, the $InCl_2$ is reduced to LiCl and $In_7Cl_9$ at a voltage below $\phi_{eq}$ = 2.26 V, as 7 $InCl_2$ + 5 Li → 5 LiCl + $In_7Cl_9$. Further, the $In_7Cl_9$ is reduced to LiCl and InCl at a voltage below $\phi_{eq}$ = 2.18 V, $In_7Cl_9$ + Li → LiCl + InCl. Furthermore, the InCl is reduced to LiCl and In at a voltage below $\phi_{eq}$ = 1.99 V, as InCl + Li → LiCl + In. The In continues reductions by Li-In alloying at lower voltages, and eventually it is reduced to $Li_{13}In_3$ in equilibrium with Li at $\phi_{eq}$ = 0.11 V. On the other hand, the oxidation reaction of LIC occurs with an electrochemical de-lithiation to $InCl_3$ and $Cl_2$ at a voltage above $\phi_{eq}$ = 4.34 V, as $Li_3InCl_6$ → 3 Li + $InCl_3$ + 3/2 $Cl_2$. As a result, the LIC has an *ab initio* electrochemical stability window between 2.47 V and 4.34 V as shown in the shaded area in **Figure 1 (c)**, where LIC is electrochemically stable. For LPSC, the full reduction and oxidation reactions and the corresponding equilibrium phases are shown in **Table S3**. The LPSC has an electrochemical stability window between 1.74 V and 2.01 V as shown in the shaded area in **Figure 1 (d)**. However, the LIC and LPSC electrolytes have much higher reduction potential than a Li-anode. This suggests that the LIC and LPSC electrolytes are thermodynamically favored to be reduced and decomposed when they work with Li-anode.

The chemical stability of LIC and LPSC with a pure Li-anode without an applied voltage can be calculated by constructing a pseudo-binary phase diagram with the end-member compositions LIC-Li or LPSC-Li in the Li-In-Cl or Li-P-S-Cl phase diagrams in **Figures 1 (a) and 1 (b)**, respectively. This methodology of the chemical stability calculation was first developed by Ceder et al., and it is intimately related to the bulk thermodynamics.[76] All possible competing phases and resulting chemical reaction energies were calculated as a function of molar fraction $x$ for Li (i.e., molar fraction $1 - x$ for the LIC in **Table S4** or LPSC in **Table S5**). For instance, a mixture of LIC and Li has a reaction energy of -517 meV/atom at a molar fraction $x = 0.769$ with respect to the competing phases (chemical reaction products) In and LiCl, as 3 $Li_3InCl_6$ + 10 Li → $LiIn_3$ + 18 LiCl. **Figure 1 (e) and 1 (f)** show pseudo-binary phase diagrams with chemical reaction energies for the end-member compositions LIC-Li and LPSC-Li as a function of the molar fraction $x$ of Li, respectively. The pseudo-binary phase diagrams are shown as a function of the atomic fraction of Li in **Figure S1**. Over the whole molar fraction, the two mixtures LIC-Li and LPSC-Li have very negative chemical reaction energy, so there is a great thermodynamic driving force for chemical reactions of LIC-Li and LPSC-Li. This indicates that LIC or LPSC are relatively thermodynamically unstable when in physical contact with Li-anode.

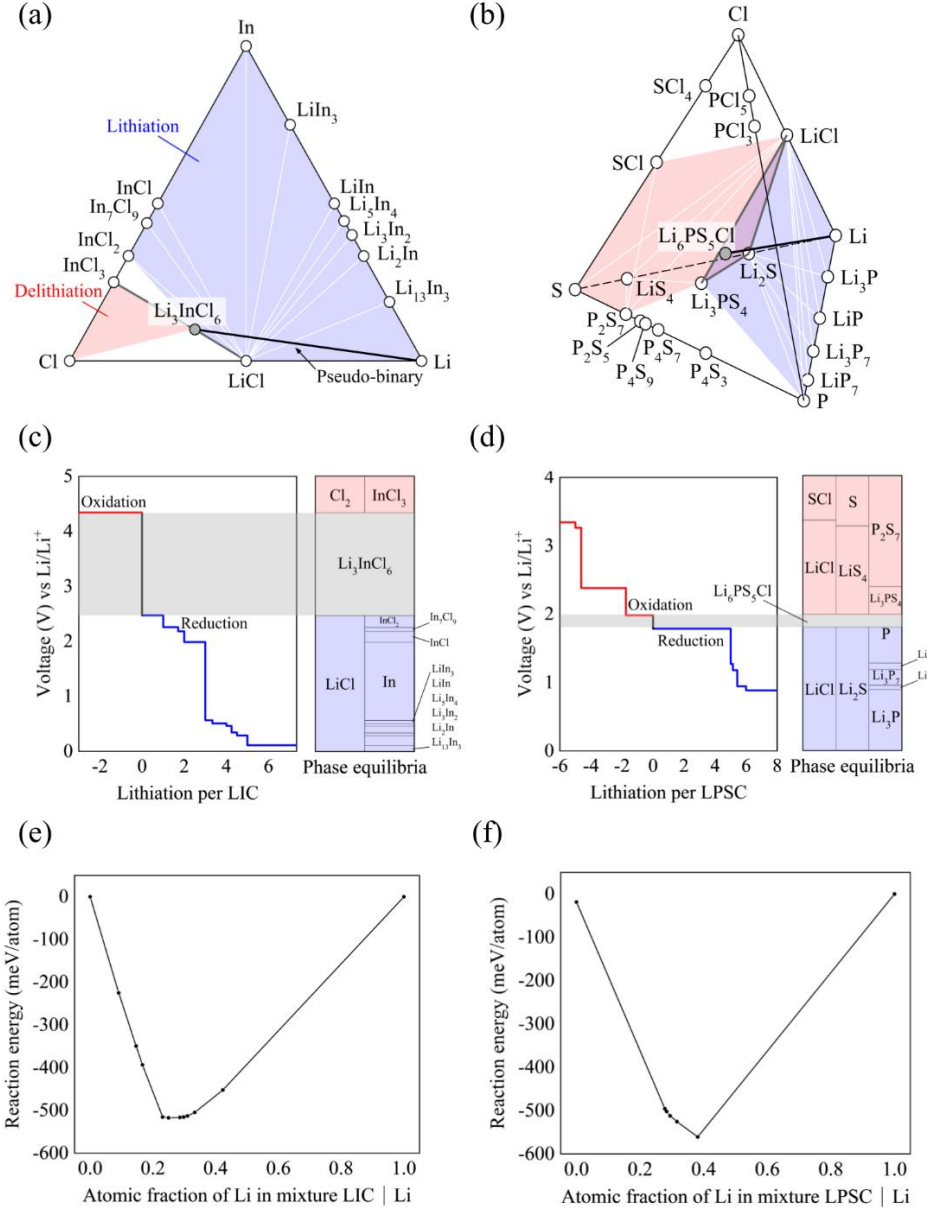

**Figure 1**. The phase diagrams of the (a) Li-In-Cl and (b) Li-P-S-Cl systems highlight the lithiation (reduction) and delithiation (oxidation) phase equilibrium regions of $Li_3InCl_6$ (LIC) and $Li_6PS_5Cl$ (LPSC), which are shaded red and blue, respectively. The equilibrium potential profiles and phase equilibria are shown for (c) $Li_3InCl_6$ (LIC) and (d) $Li_6PS_5Cl$ (LPSC). Along the tie lines (black lines) of LIC-Li and LPSC-Li in the phase diagrams, pseudo-binary phase diagrams were calculated for (e) LIC-Li and (f) LPSC-Li as a function of the atomic fraction $x$ of Li.

**Solid-electrolyte Interphase Stability**

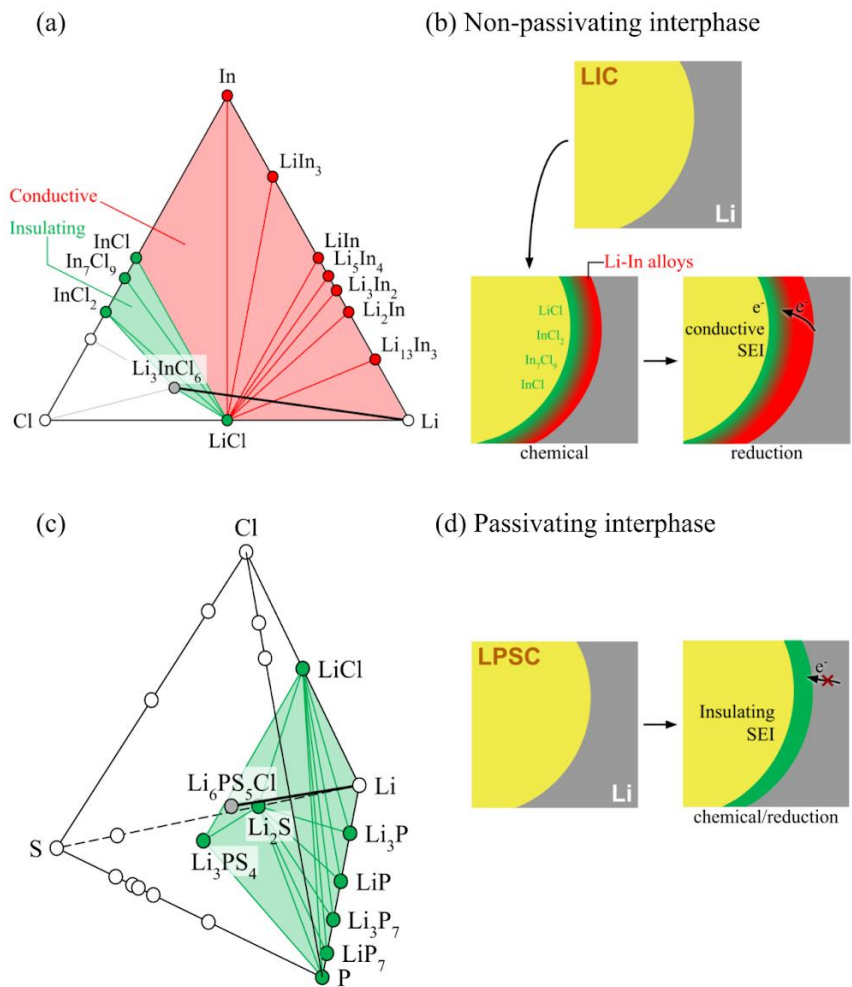

**Figure 2**. The solid-electrolyte interphase (SEI) composition regions for (a) LIC-Li, where insulating and conductive phase equilibria are colored green and red, respectively. (b) The Li-In alloys in the non-passivated SEI offer an electronically conductive network, leading to a continuous reduction reaction of LIC. The SEI compositions of (c) LPSC-Li are entirely insulating (green), and (d) the further reduction of LPSC can be kinetically passivated.

The electrochemical and chemical reactions between SSEs and a pure Li-anode would produce reaction products and form an interphase layer at the anode interface, which are commonly called solid-electrolyte interphase (SEI). If these parasitic reactions occur continuously, the SEI will grow, thereby exacerbating cycling performance. Conversely, if the SEI growth ceases, it becomes a passivated interphase layer. It has been found that the formation of an electronically conductive SEI leads to continuous electrochemical reactions of SSEs and results in the thickening of the SEI. Therefore, it is crucial to understand the reaction products formed in the SEI and their electronic conductivity. Due to the lack of electronic conductivity data, the Kohn-Sham band gaps calculated in GGA were used to indicate the electronic conductivity.

For the LIC-Li interface, all possible phases of the reduction and chemical reactions are illustrated in the colored areas of the Li-In-Cl phase diagram in **Figure 2 (a)**. The SEI composition regions for LIC-Li that are insulating and conductive are colored green and red, respectively, and their Kohn-Sham band gaps are presented in **Figure S2 (a)**. The Li-In alloys would be formed mostly through the reduction and chemical reactions between LIC and Li; however, they possess zero Kohn-Sham band gaps and exhibit electronically conductive phases formed in SEI. **Figure 2 (b)** illustrates that the distribution of Li-In alloys in the interphase layer offers an electronically conductive network, leading to continuous reduction reactions between LIC and the Li-anode. This thickens the interphase layers, thereby exacerbating the cycling performance. For the LPSC-Li interface, all possible phases formed through the reduction and chemical reactions as determined in the green areas of the Li-P-S-Cl phase diagram in **Figure 2 (c)**, which exhibits electronic insulating species with large band gaps, as shown in **Figure S2 (b)**. **Figure 2 (d)** illustrates that the electron transfer is impeded by this electronic insulating interphase, and therefore the further reduction of LPSC can be kinetically mitigated. This suggests that the as-formed interphase between LPSC and Li can act as a passivated layer.

**Protective Interlayers**

Interlayers introduced at the SSE-anode interface have been widely used to enhance anode interface stability. To the best of our knowledge, 29 ceramic materials of oxides, nitrides, sulfides and halides and 15 kinds of alloys have been reported to improve anode interface stability, as shown in **Table S6**.[20-66] An ideal anode interlayer is expected to be a stable phase with chemically and electrochemically stable interfaces with SSEs and the anode, high ionic conductivity, and low electronic conductivity. However, the performance of interlayers depends not only on the chemical properties of the material but also on the interlayer technology and processing parameters, complicating the identification of the most promising interlayer materials. To identify the best-performing anode interlayer materials, we employed *ab initio* calculations to obtain the convex hull energy of phase, electrochemical stability window, interfacial chemical reaction energy, Li-ion migration barrier, and electronic band gap of the reported interlayer materials. **Figure 3** illustrates the computational screening process for the 44 kinds of the anode interlayers, where a total of 100 compositions were screened.

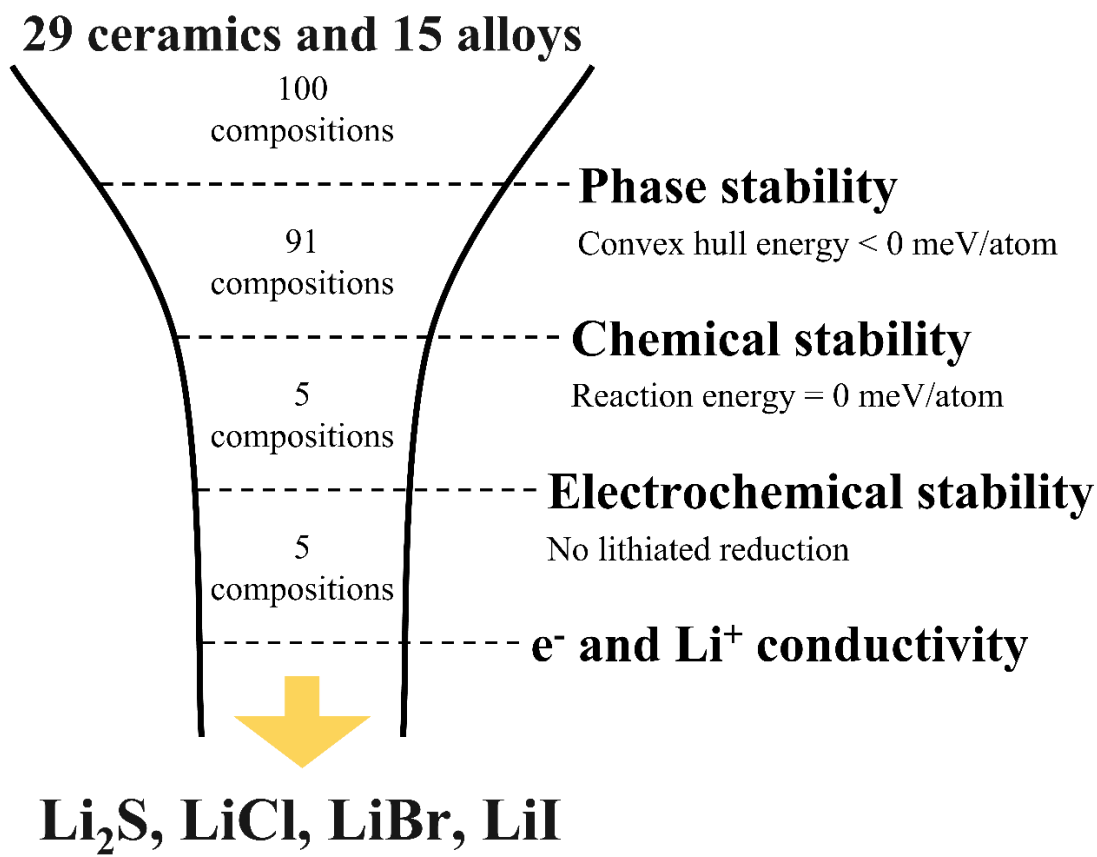

**Figure 3**. Flowchart describing *ab initio* screening of the 100 compositions from the 29 kinds of ceramics and 15 kinds of alloys reported for being effective Li-anode interlayer materials.

The almost reported interlayer materials have negative convex hull energy except for $Cu_3N$, $Li_3ScCl_6$, $LiMg_2$, $Li_3Mg$, $LiIn_3$, $LiIn$, $Li_3Sn$, $Zn_3Cu$ and $ZnCu$ in **Table S6**, and they are thermodynamically stable phases. The interlayer provides two new interfaces of SSE-interlayer and interlayer-anode, which should have higher interfacial chemical stability than the original SSE-anode interface. For example, $Al_2O_3$ shows less negative chemical reaction energy and higher chemical stability with Li, LPSC, and LIC in **Figure S3**, respectively. Nevertheless, the mutual chemical reaction energy varies with the mixing fraction as shown in the pseudo-binary phase diagram. A minimum chemical reaction energy, as indicated by the symbol stars in **Figure S3** is defined as interfacial chemical reaction energy for quantifying the interfacial chemical stability of SSE-interlayer or interlayer-Li interfaces. **Figure 4** shows that most of the reported interlayer materials still have negative interfacial chemical reaction energy with Li, LPSC or LIC. The 6 ceramic materials $La_2O_3$, $Li_3N$, $Li_2S$, LiF, LiCl, LiBr, and LiI, and the 15 alloy compositions $Li_5Mg$, $Li_2Al$, $Li_{21}Si_5$, LiZn, $Li_{15}Ge_4$, $Sr_{19}Li_{44}$, $Li_3Ag$, $Li_{13}In_3$, $Li_{17}Sn_4$, $Li_{15}Au_4$, $Li_3Hg$, $LiCu_3$, $Li_2ZnCu_3$, $Li_3Bi$, and $LiC_{12}$ have no thermodynamically favorable chemical reaction with Li, and they are stable with Li-anode. Among the 21 interlayer compositions screened, LiF and LiCl are stable with LIC, while $Li_2S$, LiF, LiCl, LiBr and LiI are stable with LPSC. Especially, only LiF and LiCl are stable with both LPSC and LIC. As a result, the five interlayers $Li_2S$, LiF, LiCl, LiBr and LiI are selected for further evaluation of their electrochemical stability, electronic conductivity, and ionic conductivity.

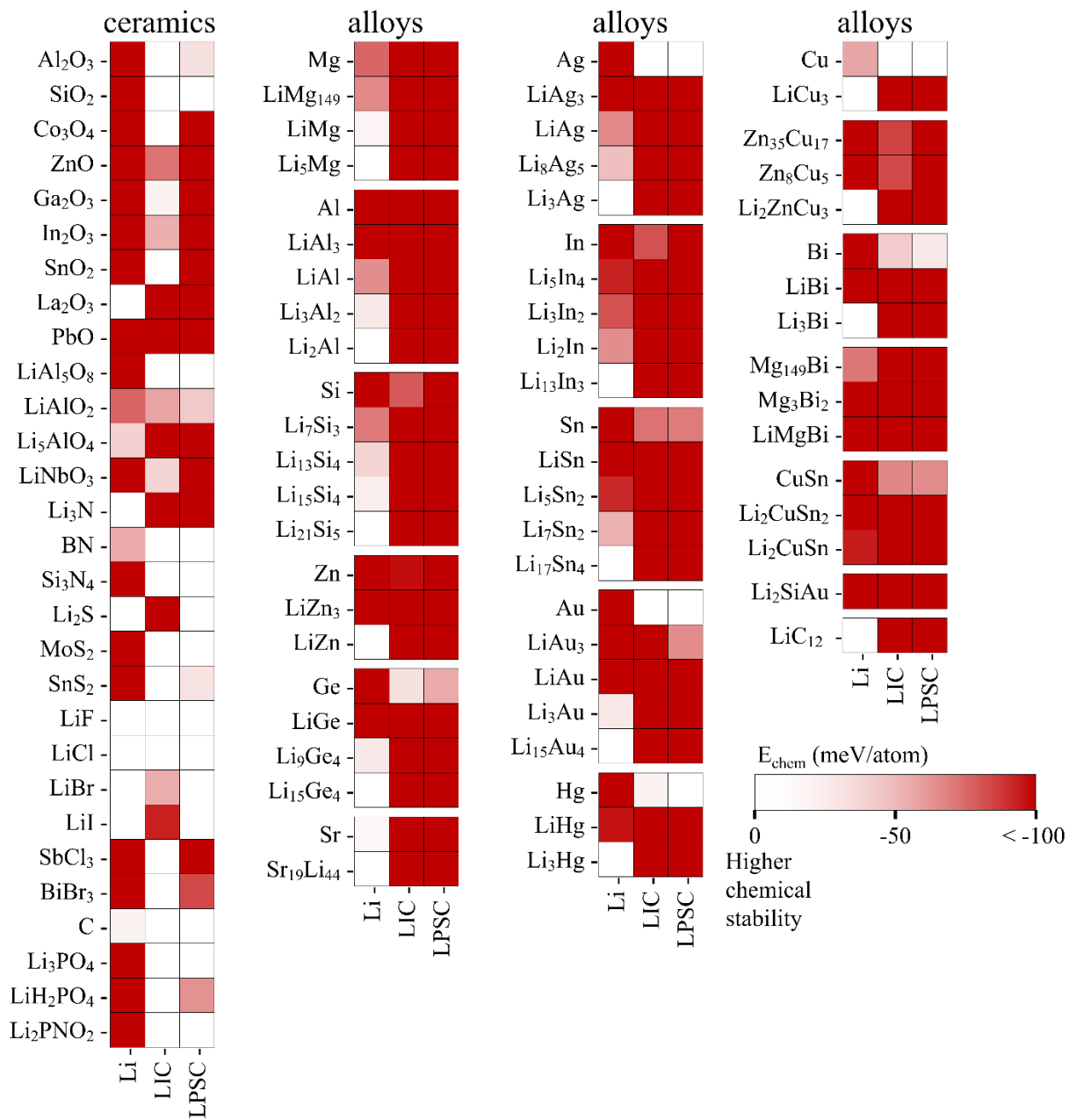

**Figure 4**. Interfacial chemical reaction energy of the ceramics- and alloy-interlayers with Li, LIC, and LPSC. The color of each square in the heatmap indicates the minimum chemical reaction energy of the interlayer-Li, interlayer-LIC and interlayer-LPSC. White color denotes high chemical stability, while dark red color denotes low chemical stability.

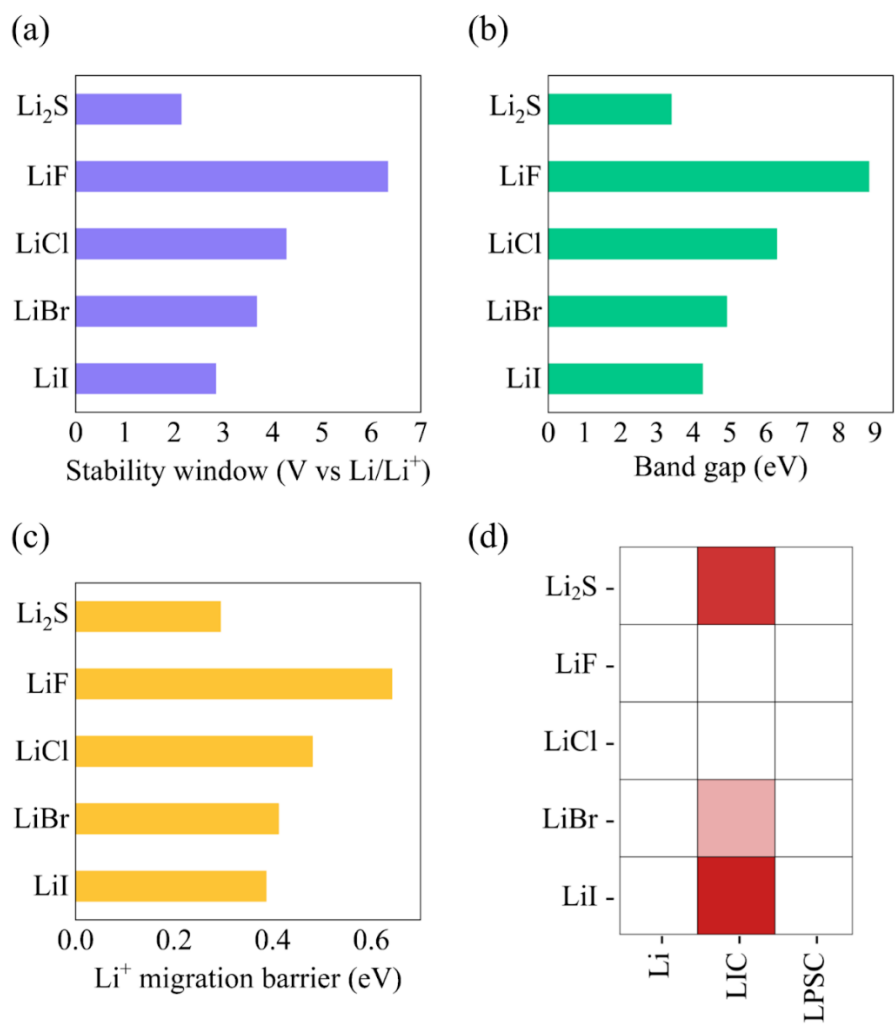

**Figure 5**. (a) Electrochemical stability window, (b) band gap, (c) NEB-determined Li+ migration barrier of the candidate interlayers Li$_2$S, LiF, LiCl, LiBr, and LiI, and (d) heatmap showing the minimum chemical reaction energy of the five interlayers against Li, LIC, and LPSC.

**Figure 5 (a)** shows the ab initio electrochemical stability window for Li$_2$S, LiF, LiCl, LiBr, and LiI. The five interlayers have no electrochemical reduction ability, so they are all electrochemically stable at 0 V versus Li-anode. **Figure 5 (b)** shows the five interlayer materials have high band gaps (> 3 eV), indicating they can impede electron transport and kinetically suppress the further electrochemical reduction of electrolytes. Moreover, interlayers are not expected to hinder the transport of Li ions, and ideal interlayer materials should have a high enough Li diffusivity. **Figure 5 (c)** shows the Li migration barriers for the five candidate interlayer materials, calculated using the CI-NEB method, as shown in **Figure S4**. However, LiF has a high Li migration barrier of 0.64 eV, which is consistent with the experimentally measured low ionic conductivity of $5 \times 10^{-10} - 3 \times 10^{-9}$ Scm$^{-1}$.[79,80] Li$_2$S, LiCl, LiBr, and LiI show low Li migration barriers of 0.29, 0.48, 0.41, and 0.39 eV, respectively, which are consistent with the experimentally measured ionic conductivity of ~$10^{-5}$ Scm$^{-1}$, ~$10^{-9}$ Scm$^{-1}$, ~$10^{-9}$ Scm$^{-1}$ and ~$10^{-7}$ Scm$^{-1}$.[80-83] Among the five interlayer candidates, Li$_2$S, LiCl, LiBr, and LiI show high reduction stability versus Li-anode, low electronic conductivity, and high enough Li-ion conductivity. By further considering the chemical stability with Li, LIC and LPSC in **Figure 5 (d)**, all Li$_2$S, LiCl, LiBr, and LiI are stable with Li and LPSC; however, for interfacing with LIC, only LiCl is stable, and Li$_2$S, LiBr and LiI are unstable. Overall, LiCl could be the best performing interlayer material for LIC-Li, and Li$_2$S and LiI could be the best ones for LPSC-Li in ASSLBs.

**Conclusions**

*Ab initio* calculation results show that the two emerging, promising halide LIC and argyrodite LPSC electrolytes are not thermodynamically chemically and electrochemically compatible with a Li-anode in ASSLBs. Surprisingly, the insulating solid-electrolyte interphase (SEI) of LPSC-Li could kinetically impede the further reduction of LPSC; however, the electronic conductive SEI of LIC-Li allows the continuous reduction of LIC. To stabilize the LIC-Li and LPSC-Li interfaces, the 44 kinds and 100 compositions of reported effective ceramics- and alloy-interlayer materials were evaluated. However, most of the reported interlayer materials are still not thermodynamically chemical stable with Li, LIC or LPSC. The Li$_2$S, LiCl, LiBr and LiI interlayer materials show high chemical and reduction stability with Li and LIC or LPSC, low electronic conductivity and high enough Li-ion conductivity. By further considering chemical stability with LIC and LPSC, LiCl could be the best performing interlayer material for LIC-Li, and $Li_2S$ and LiI could be the best ones for LPSC-Li in ASSLBs.

**Acknowledgement**

This study is financially supported by the National Science and Technology Council (NSTC) (Project Nos. 112-2221-E-011-041-MY3, 112-2923-E-006-004 and 110-2222-E-011-006-MY3), the Ministry of Education of Taiwan (MOE "Sustainable Electrochemical Energy Development Center" (SEED) project, as well as the supporting facilities from National Center for High-performance Computing (NCHC).